\documentstyle[11pt]{article}

\begin{document}

\title{Development of a Kohn-Sham like potential in the
Self-Consistent Atomic Deformation Model}

\author{M. J. Mehl, L. L. Boyer and H. T. Stokes\\
{\em Complex Systems Theory Branch}\\
{\em Naval Research Laboratory} \\
{\em Washington, D.C. 20375-5345}
}

\maketitle

\begin{abstract}
This is a brief description of how to derive the local ``atomic''
potentials from the Self-Consistent Atomic Deformation (SCAD) model
density function.  Particular attention is paid to the spherically
averaged case.
\end{abstract}

\section{The SCAD Model}
\label{sec:scadintro}

The Self-Consistent Atomic Deformation (SCAD) model \cite{boyer93}
has been described as an extension of the work of Gordon and Kim \cite
{gordo72}.  In can also be viewed as an approximation to density 
functional theory, which is the point of view taken in this paper.

The SCAD model approximates the exact electronic density functional
\cite{hohen64} by the form:
\begin{equation}
E[n({\bf r})] = \sum_i T_0[v_i({\bf r})] + \left\{T_{TF}[n({\bf r})]
- \sum_i T_{TF}[n_i ({\bf r})]\right\} + F[n ({\bf r})] ~ .
\label{equ:scdf}
\end{equation}
Here:

\begin{enumerate}
\item $T_0[n_i ({\bf r})]$ is the kinetic energy of non-interacting
electrons centered about the site ${\bf R}_i$,
\begin{equation}
T_0[n_i ({\bf r})] = - \sum_\alpha \int \psi^*_{i\alpha} ({\bf r})
\nabla^2 \psi_{i\alpha} ({\bf r}) \, d^3r ~ ,
\label{equ:kin}
\end{equation}
where the $\psi_{i\alpha}$ are an orthonormal set of wave functions
such that
\begin{equation}
n_i ({\bf r}) = \sum_\alpha |\psi_{i\alpha}({\bf r})|^2 ~ .
\label{equ:locden}
\end{equation}
(We use atomic units, with  $\hbar = 2 m = e^2 / 2
= 1$, where $m$ is the mass of the electron and $-e$ is its charge.)

\item We assume that the charge associated with each site,
\begin{equation}
N_i = \int n_i ({\bf r}) \, d^3r ~ ,
\end{equation}
is fixed.  That is, there is no charge transfer between sites during
the iteration of the problem to self-consistency.

\item The total density of the system is given by
\begin{equation}
n ({\bf r}) = \sum_i n_i ({\bf r + R}_i) ~ .
\label{equ:dendef}
\end{equation}

\item The Thomas-Fermi kinetic energy is given (in Rydberg units) by
\begin{equation}
T_{TF}[n ({\bf r})] = \frac35 (3 \pi^2)^{\frac23} \int n^{\frac53}
({\bf r}) \, d^3r ~ .
\end{equation}
The sum over atomic Thomas-Fermi kinetic energies in
(\ref{equ:scdf}) is to cancel this contribution to $E[n]$ from the
individual atoms, where the kinetic energy is given by
(\ref{equ:kin}).  The Thomas-Fermi terms in parenthesis in
(\ref{equ:scdf}) thus represent the overlap kinetic energy, and
vanish when the atoms are separated by large distances.

\item The functional $F[n ({\bf r})]$ contains the remaining density
functional terms\cite{hohen64,kohn65}, including the Coulomb
interaction terms (electron-electron, electron-nucleus, and
nucleus-nucleus), which is a simple functional of the total density
(\ref{equ:dendef}), and the exchange-correlation term, which is
local functional of the density within the Local Density
Approximation (LDA).  The functional derivative of $F$ with respect
to the total density is just the Kohn-Sham potential \cite{kohn65},
\begin{equation}
v_{KS} [n({\bf r});{\bf r}] = \frac{\delta F [n({\bf r})]}{\delta n
({\bf r})} ~ .
\end{equation}
This potential includes both the Coulomb and the
exchange-correlation potentials.
\end{enumerate}

\section{Finding the potential}
\label{sec:pot}

In this section we wish to find a way to determine the individual
charge densities $n_i ({\bf r})$.  We begin by noting that if $n_i$
is, as we assume, ``v-representable'', then there is a one-to-one
correspondence between the density $n_i$ and a potential $v_i$, such
that the wave functions described in (\ref{equ:kin}) and
(\ref{equ:locden}) satisfy the Schr\"{o}dinger equation
\begin{equation}
- \nabla^2 \psi_{i\alpha} ({\bf r}) + v_i({\bf r}) \psi_{i\alpha}
({\bf r}) = \epsilon_{i\alpha} \psi_{i\alpha} ({\bf r}) ~ ,
\label{equ:schro}
\end{equation}
This is equivalent to rewriting the SCAD density Functional
(\ref{equ:scdf}) in the form
\begin{equation}
E[\{v_i(r)\}] = \sum_i T_0[v_i(r)] + \left\{T_{TF}[n({\bf r})] -
\sum_i T_{TF}[n_i ({\bf r})]\right\} + F[n ({\bf r})] ~ .
\label{equ:sscdf}
\end{equation}

Now consider changing the potential in (\ref{equ:sscdf}) 
\begin{equation}
v_i({\bf r}) \rightarrow v_i({\bf r}) + \delta v_i({\bf r}) ~ .
\end{equation}
This change in the potential at the site $i$ changes the associated
wave functions
\begin{equation}
\psi_{i\alpha} ({\bf r}) \rightarrow \psi_{i\alpha} ({\bf r}) +
\delta \psi_{i\alpha} ({\bf r}) ~ ,
\end{equation}
where, to first order in $\delta v_i$,
\begin{equation}
[\epsilon_{i\alpha} + \nabla^2 - v_i ({\bf r})] \delta \psi_{i\alpha} ({\bf
r}) = (\delta v_i (r) - \delta \epsilon_{i\alpha}) \psi_{i\alpha}
({\bf r}) ~ .
\label{equ:del}
\end{equation}
Without loss of generality we may take each $\delta \psi$ to be
orthogonal to the corresponding $\psi$:
\begin{equation}
\int \delta \psi_{i\alpha}^* ({\bf r}) ~ \psi_{i\alpha} ({\bf r}) =
0.
\label{equ:dorth}
\end{equation}
Multiplying both sides of (\ref{equ:del}) by $\psi_{i\alpha}^* ({\bf
r})$ and integrating, we obtain the first order change in the
eigenvalue:
\begin{equation}
\delta \epsilon_{i\alpha} = \int |\psi_{i\alpha} ({\bf r})|^2 \delta
v_i ({\bf r}) \, d^3r ~ .
\label{equ:deleps}
\end{equation}

The electron density on site $i$ then changes by an amount
\begin{equation}
\delta n_i ({\bf r}) = 2 \Re \sum_\alpha \delta \psi_{i\alpha}^* ({\bf r})
~ \psi_{i\alpha} ({\bf r}) ~ ,
\label{equ:delden}
\end{equation}
which is a change of $O[\delta v_i]$.  ($\Re z$ is the real part of
$z$.)  Note that the densities $n_j$ on the other sites will not
change, since by supposition we are only changing the spherical
potential on site $i$, and there is no charge transfer.  Thus
\begin{equation}
\delta n ({\bf r}) = \delta n_i ({\bf r + R}_i) ~ .
\end{equation}

The kinetic energy $T_0$ (\ref{equ:kin}) changes by an amount
\begin{equation}
\delta T_0 = - 2 \Re \sum_\alpha \int \delta \psi^*_{i\alpha} ({\bf r})
\nabla^2 \psi_{i\alpha} ({\bf r}) \, d^3r ~ .
\end{equation}
By (\ref{equ:schro}) this becomes
\begin{equation}
\delta T_0 = 2 \Re \sum_\alpha \int \delta\psi_{i\alpha} ({\bf r})
[\epsilon_{i\alpha} - v_i ({\bf r})] \, \psi_{i\alpha} ({\bf r}) \,
d^3r ~ .
\end{equation}
Applying (\ref{equ:dorth}) and (\ref{equ:delden}), we find
\begin{equation}
\delta T_0 = - \int v_i ({\bf r}) \, \delta n_i ({\bf r}) ~ .
\end{equation}

The other parts of the density functional change in a
straight-forward way:
\begin{equation}
\delta F[n ({\bf r})] = \int v_{KS} [n({\bf r});{\bf r}] \, \delta
n_i ({\bf r}) \, d^3r ~ ,
\end{equation}
\begin{equation}
\delta T_{TF} [n({\bf r})] = \int v_{TF} [n({\bf r});{\bf r}] \,
\delta n_i ({\bf r}) \, d^3r ~ ,
\end{equation}
and
\begin{equation}
\delta T_{TF}[n_i({\bf r})] = \int v_{TF} [n({\bf r});{\bf r}] \,
\delta n_i ({\bf r}) \, d^3r ~ ,
\end{equation}
where
\begin{equation}
v_{TF}[n ({\bf r}),{\bf r}] = \frac{\delta T_{TF} [n({\bf
r})]}{\delta n ({\bf r})} = (3 \pi^2)^{\frac23} n^{\frac23}
({\bf r})
\end{equation}
is the Thomas-Fermi potential.

Substituting equations (19-23) into the energy formula
(\ref{equ:sscdf}), we find that it changes by an amount
\begin{equation}
\delta E[\{ v_i (r) \}] = \int \left\{ v_{KS}[n({\bf r});{\bf r}] +
v_{TF}[n({\bf r});{\bf r}] - v_{TF}[n_i({\bf r});{\bf r}] - v_i({\bf
r}) \right\} \, \delta n_i ({\bf r}) \, d^3r ~ .
\label{equ:dele}
\end{equation}
Equation (\ref{equ:dele}) must hold for an arbitrary infinitesimal
change in the potential $\delta v_i$.  By (\ref{equ:delden}) and
(\ref{equ:del}), this produces a corresponding change in $\delta
n_i$.  Hence (\ref{equ:dele}) must hold for arbitrary infinitesimal
changes in $\delta n_i$, which can only occur if the term in curly
brackets vanishes, i.e., if
\begin{equation}
v_i ({\bf r}) = v_{KS}[n({\bf r});{\bf r}] + v_{TF}[n({\bf
r});{\bf r}] - v_{TF}[n_i({\bf r});{\bf r}] ~ .
\label{equ:videf}
\end{equation}

\section{Spherical Averages}
\label{sec:scadavg}

When using the SCAD model for crystals with atoms at high symmetry sites,
 it is often useful to
assume that the densities $n_i$ are spherically symmetric.  This can
be achieved by assuming that the $n_i$ represent a closed shell
system, and that each potential $v_i$ is spherical.  Under these
assumptions, the formalism developed in Section~\ref{sec:pot}
remains the same until equation (\ref{equ:dele}), except that we
replace the arbitrary potential $v_i ({\bf r})$ by the spherical
potential $v_i (r)$, and restrict the potential changes to the
spherical $\delta v_i (r)$.  At that point we note that because of
the assumption that $n_i$ represents a closed shell, $\delta n_i
({\bf r})$ must also be spherical.  Thus (\ref{equ:dele}) can be
written
\begin{equation}
\delta E[\{ v_i (r) \}] = \int \left\{ v_{KS}[n({\bf r});{\bf r}] +
v_{TF}[n({\bf r});{\bf r}] - v_{TF}[n_i({\bf r});{\bf r}] - v_i(r)
\right\} \, \delta n_i (r) \, d^3r ~ .
\label{equ:deles}
\end{equation}
This holds for arbitrary spherical potential changes $\delta v_i
(r)$, whence $v_i (r)$ must obey
\begin{equation}
v_i (r) = \frac{1}{4 \pi} \int \left\{ v_{KS}[n({\bf r});{\bf r}] +
v_{TF}[n({\bf r});{\bf r}] - v_{TF}[n_i({\bf r});{\bf r}] \right\}
\, d\Omega ~ ,
\end{equation}
where the integral is over the surface of a sphere of radius r.
Note that this is just the spherical average of the full potential
(\ref{equ:videf}):
\begin{equation}
v_i (r) = \frac{1}{4 \pi} \int v_i ({\bf r}) \, d\Omega ~ .
\end{equation}

\section{Discussion}
\label{sec:sscad}

The essential difference between the SCAD model and the more familiar
Kohn-Sham LDA formulation is that the SCAD model approximates kinetic
energy in low-density regions (outside atomic cores) with the aid of a
local functional.  While the total density in the SCAD model is expressed 
as a sum of localized densities, this should not be a major source of 
error, provided that the localized densities are given adequate flexibility.
Solving the atomic calculations self consistently, with the potential
for each site determined from the densities and potentials of the rest
of the system, puts the atomic energy levels on a common scale.  Just
as in the Kohn-Sham treatment, the lowest energy levels are occupied 
to achieve the minimum total energy.  

The SCAD model offers some advantages over the Kohn-Sham LDA approach.
It is generally more efficient, particularly for large systems, because
the computational labor increases linearly with the number of atoms in
the system.  Moreover, it is ideally suited for application on a 
massively parallel computer.  Finally, we note that the SCAD model
has the correct behavior in the limit of infinitely separated atoms,
whereas, each electron in the Kohn-Sham model is delocalized
and remains so in this limit. 

\section{Acknowledgment}
\label{sec:ack}

This work was supported by the Office of Naval Research.


\begin{thebibliography}{99}
\bibitem{boyer93}L. L. Boyer and M. J. Mehl, {\em Ferroelectrics},
{\bf 150}, 13-24 (1993).

\bibitem{gordo72}R. G. Gordon and Y. S. Kim, {\em J. Chem. Phys.} {\bf
56}, 3122 (1972).

\bibitem{hohen64}P. Hohenberg and W. Kohn, {\em Phys. Rev.} {\bf
136}, B864 (1964).

\bibitem{kohn65}W. Kohn and L. J. Sham, {\em Phys. Rev.} {\bf 140},
A1133 (1965).
\end{thebibliography}
\end{document}